\documentclass[aps,prl,reprint,groupedaddress]{revtex4-1}
\usepackage{amsmath}

\newcommand{\beq}{\begin{equation}}
\newcommand{\eeq}{\end{equation}}

\begin{document}

\title{Near-horizon Carroll symmetry and black hole Love numbers}
\author{Robert F. Penna}
\email[]{rp2835@columbia.edu}
\affiliation{Department of Physics, Columbia University, 538 West 120th Street, New York, NY 10027}

\begin{abstract}

According to the black hole membrane paradigm, the black hole event horizon behaves like a 2+1 dimensional fluid.  The fluid has nonzero momentum density but zero velocity.  As a result, it does not respond to tidal forces in the usual way.   In this note, we point out that this unusual behavior can be traced back to an emergent, near-horizon Carroll symmetry (the Carroll group is the $c\rightarrow 0$ limit of the Poincar\'e group).  For Schwarzschild black holes in $d=4$ general relativity, we relate the vanishing of the black hole fluid's velocity to vanishing of the black hole's Love numbers.  This suggests near-horizon Carroll symmetry may have a role to play in explaining black hole Love numbers.

\end{abstract}

\pacs{}

\maketitle

\section{Introduction}

The new era of gravitational wave astronomy offers opportunities to test fundamental physics.  LIGO observations of neutron star mergers can probe the neutron star equation of state.  Observations of black hole mergers can test the Kerr model of spinning black holes and search for physics beyond general relativity.  In either case, it is crucial to understand how a body in a binary responds to tidal forces from its companion.  

During the inspiral phase of a merger, when the bodies are widely separated, the problem can be studied in perturbation theory.  In this case, the response of a body to external tidal forces can be characterized by a set of parameters called Love numbers. The first Love number is defined by considering the response of a body's quadrupole moment, $Q_{ij}$, to an external quadrupolar tidal field, $\mathcal{E}_{ij}$.  One expects
\beq
Q_{ij} \propto \mathcal{E}_{ij}.
\eeq
The constant of proportionality (suitably normalized) \cite{Binnington:2009bb} is the first Love number.  It is the first in an infinite sequence of Love numbers, one for each multipole moment.  In general relativity, there are in fact two infinite sets of Love numbers.  They correspond to the tidal field's gravitoelectric and gravitomagnetic components.  The Love numbers of neutron stars depend on the neutron star equation of state, so Love number measurements probe the equation of state.  It turns out the Love numbers of Schwarzschild black holes are all zero (at least in asymptotically flat, $d=4$ general relativity) \cite{Binnington:2009bb,Damour:2009vw,Kol:2011vg,Chakrabarti:2013lua,Chakrabarti:2013xza}.  So black hole Love number measurements are tests of general relativity.

The fact that Love numbers are zero for $d=4$ Schwarzschild black holes is surprising from an effective field theory (EFT) perspective and raises a puzzle \cite{unpublished,Rothstein:2014sra,Porto:2016zng}.    In the EFT for binary inspiral \cite{Goldberger:2004jt,Goldberger:2005cd,Goldberger:2006bd,Goldberger:2007hy,Porto:2016pyg,Rothstein:2003mp,Rothstein:2014sra}, the merging bodies are treated as point particles and the Love numbers enter as couplings.  
From the EFT perspective, one does not expect the Love numbers to be zero unless a symmetry forces them to vanish.  Absent a symmetry explanation, the theory with vanishing Love numbers looks fine tuned.  This could be a hint that the EFT is incomplete; perhaps new physics will enter and eliminate the tuning.  Rothstein and Goldberger \cite{unpublished} (see also \cite{Rothstein:2014sra,Porto:2016zng}) have noted that the binary inspiral EFT, as presently understood, appears to be tuned in this sense. 

In this note, we reconsider the description of black hole tidal interactions using the membrane paradigm and confront a similar puzzle.  According to the membrane paradigm, the black hole horizon behaves like a $2+1$ dimensional fluid.  The fluid has nonzero momentum density but but zero velocity.  As a result, it does not respond to tidal forces in the usual way \cite{Price:1986yy}.  We observe that vanishing of the black hole fluid velocity follows from an emergent, local Carroll symmetry in the near-horizon region.  The Carroll group is the $c\rightarrow 0$ limit of the Poincar\'e group \cite{levy1965nouvelle,Duval:2014lpa,Duval:2014uoa,Duval:2014uva,
Bergshoeff:2014jla,
Bagchi:2016bcd,
Bergshoeff:2017btm,
Duval:2017els,
Ciambelli:2018xat,
Ciambelli:2018wre,
Zhang:2018gzn,
Barducci:2018wuj,
Ciambelli:2018ojf}.
The Carroll group is a subgroup of the full Bondi-van der Burg-Metzner-Sachs (BMS) group \cite{Bondi:1962px,Sachs:1962wk,Sachs:1962zza} and it is also interesting to contemplate whether a near-horizon version of the full BMS group \cite{Blau:2015nee,Donnay:2015abr,Afshar:2016wfy,Donnay:2016ejv,Eling:2016xlx,Hawking:2016msc,Mao:2016pwq,Lust:2017gez,Chandrasekaran:2018aop,Donnay:2018ckb,Penna:2015gza,Penna:2017bdn} may have a signature in black hole tidal interactions.  We leave this possibility for the future.

For Schwarzschild black holes in $d=4$ general relativity, we relate the vanishing of the membrane paradigm fluid velocity to vanishing of the black hole's Love numbers.  This suggests Carroll symmetry may have a role to play in explaining black hole Love numbers.  However, black hole Love numbers are nonzero in higher spacetime dimensions \cite{Kol:2011vg}, in asymptotically anti de Sitter spacetimes \cite{Emparan:2017qxd}, and in modified theories of gravity \cite{Cardoso:2018ptl}.  So Carroll symmetry alone cannot be the full story.  Still we believe the observations in this note might hold a clue for future investigations.

\section{Result}

We begin by reviewing a surprising feature of black holes first pointed out long ago by Price and Thorne \cite{Price:1986yy}.  They were studying black hole tidal interactions using the black hole membrane paradigm, which is a way of reformulating black hole dynamics in terms of a 2+1-dimensional fluid membrane living near the event horizon.  The membrane paradigm relates problems in black hole physics to possibly more familiar problems in fluid dynamics.  In retrospect, the membrane paradigm was a precursor of AdS/CFT and modern holographic theories.  Let us review what Price and Thorne found.

To begin, recall how an ordinary fluid responds to an applied tidal field.  For simplicity, consider a pressureless, inviscid fluid in two space dimensions with spatial velocity vector, $v$, in a background gravitational potential, $\Phi$.  In this case, the dynamics is governed by the Euler equation,
\beq\label{eq:euler}
\partial_t v + v\cdot \nabla v = -\nabla \Phi.
\eeq
Taking a gradient, and then extracting the symmetric trace-free (STF) part, gives the tidal force equation:
\beq\label{eq:tidalforce}
\partial_t \sigma_{ij} + v\cdot \nabla \sigma_{ij} + \theta \sigma_{ij} = -\mathcal{E}_{ij}.
\eeq
Here $\sigma_{ij}=v_{i;j}^{{\rm STF}}$ is the shear tensor, $\theta=\nabla \cdot v$ is the expansion scalar, and $\mathcal{E}_{ij}=\Phi_{;ij}^{{\rm STF}}$ is the tidal field tensor.  In Cartesian coordinates, $\sigma_{ij}=v_{(i,j)}-\frac{1}{2}\delta_{ij}{v^k}_{,k}$.  The fluid's shear tensor responds to the applied tidal field according to \eqref{eq:tidalforce}.

Now according to the black hole membrane paradigm, the black hole event horizon behaves like a 2+1 dimensional fluid.  
It has a momentum density that evolves according to a version of the Navier-Stokes equation.  Therefore one might hope to find a tidal force equation analogous to \eqref{eq:tidalforce} governing black hole tidal interactions.
So it came as something of a surprise when Price and Thorne found that the shear of the black hole membrane measured by a local observer actually vanishes.  In particular, the membrane's shear is completely independent of the applied tidal field, $\mathcal{E}_{ij}$. 

To understand this result, recall the definition of the membrane's momentum density:
\beq
\Pi_a^H \equiv {S^{\hat{0}}}_a,
\eeq
where $S_{ab}$ is the membrane's Brown-York stress energy tensor, defined in terms of the membrane's extrinsic curvature as $S_{ab}=\frac{1}{8\pi}(K_{ab}-{K^c}_c h_{ab})$.  Indices $a,b,c\dots$ run over the $2+1$ coordinates on the membrane and $h_{ab}$ is the induced metric on the membrane.  The hatted index, $\hat{0}$, indicates the timelike-direction of the local observer's reference frame.  

Just as for an ordinary fluid, the momentum density is a product, 
\beq
\Pi_a^H = (\text{\emph{energy density}})\allowbreak\times (\text{\emph{velocity}}).
\eeq
The locally measured energy density is infinite at the event horizon because a local observer hovering at the horizon is infinitely accelerated \footnote{Compare to the well-known fact that the local temperature is infinite at the horizon of Rindler space.}.  To regulate the divergent local energy density at the horizon, it is standard to introduce a timelike ``stretched horizon'' just outside the event horizon.  Let $\alpha_H\ll 1$ be the lapse function on the stretched horizon (at the event horizon, $\alpha_H\rightarrow 0$).   The local energy density on the stretched horizon is $\alpha_H^{-1}\kappa/(8\pi)$, where $\kappa$ is the black hole's surface gravity. For a weakly perturbed hole, the momentum density is \cite{Thorne:1986iy}
\beq
\Pi^H_a \approx 
	\frac{1}{8\pi}\alpha_H^{-1} \kappa v_a,
\eeq
where $v_a$ is the velocity of the membrane fluid.  The key point is that that the velocity is an extremely small, $O(\alpha_H)$, quantity.  It balances the huge, $O(\alpha_H^{-1})$, locally measured energy density to give a finite momentum density as $\alpha_H\rightarrow 0$.  Now $v_a=O(\alpha_H)$ implies that the shear tensor,  $\sigma_{ab}=v_{(a,b)}-\frac{1}{2}\delta_{ab}{v^c}_{,c}$, also vanishes at the event horizon.  In particular, the locally measured shear is completely independent of external tidal forces \footnote{Observers at infinity can measure nonzero horizon shear.  However, reference frames near the horizon are themselves unavoidably sheared by tidal fields and find vanishing net shear (for further discussion, see \cite{Price:1986yy}).}.  Price and Thorne were careful to stress this counterintuitive difference between ordinary fluids and the black hole fluid in their work.  

This is reminiscent of the story for black hole Love numbers.   In both cases, the black hole fails to respond to applied tidal fields in somewhat surprising fashion.  Below we exhibit a direct relationship between the two phenomena for $d=4$ Schwarzschild black holes.  But first we wish to describe the underlying symmetry that forces the black hole fluid velocity to vanish.  This observation is new and forms one of the main results of the present note.

\subsection{Carroll symmetry}

The Carroll group is the $c\rightarrow 0$ limit of the Poincar\'e group.  Rotations and translations act the same, but Carroll boosts act as
\begin{align}
\mathbf{x}' &= \mathbf{x},\\
t' &=t-\mathbf{b}\cdot \mathbf{x},
\end{align}
where $\mathbf{b}$ is the Carroll boost parameter.  In two spacetime dimensions the Carroll group is dual to the Galilei group (the $c\rightarrow \infty$ limit of Poincar\'e), with the roles of space and time interchanged.

Carroll-invariant particles cannot move.  Intuitively, this is obvious: we have taken $c\rightarrow 0$, and physical speeds are bounded by $c$.  (For a proof using coadjoint orbits, see \cite{Duval:2014uoa}).  We will now show that there is an emergent local Carroll symmetry at the event horizon.  This provides a symmetry explanation for Price and Thorne's observation that the black hole fluid has vanishing velocity (and, as we will see, Love numbers).

First, define Carroll space as the $c\rightarrow 0$ limit of Minkowski space:
\beq\label{eq:carroll}
ds^2 = 0\cdot dt^2 + dx^2 + dy^2 + dz^2.
\eeq
The Carroll group can be equivalently defined as the group of symmetries that preserve the Carroll metric \eqref{eq:carroll} and the Carroll connection $\Gamma^i_{jk}=0$ and commute with $\partial_t$ \cite{Duval:2014uoa}.  Let us explain these requirements in more detail.  The group of symmetries preserving the metric \eqref{eq:carroll} alone is infinite dimensional.  It includes ordinary isometries of the spatial part of the metric as well as an infinite dimensional group of ``supertranslations,'' $t\rightarrow t+ f(t,x,y,z)$.  Demanding that the symmetries preserve the connection and commute with $\partial_t$ cuts down this infinite dimensional group to the finite dimensional Carroll group defined earlier as the $c\rightarrow 0$ limit of Poincar\'e.  To see this, note that the Lie derivative of a connection is 
\begin{align}
(\mathcal{L}_X \Gamma)^i_{jk}
	= &X^p \partial_p \Gamma^i_{jk}
		+\partial_j \partial_k X^i
		-\Gamma^p_{jk}\partial_p X^i\notag\\
		&+\Gamma^i_{pk}\partial_j X^p
		+\Gamma^i_{jp}\partial_k X^p.
\end{align}
The Carroll connection is $\Gamma^i_{jk}=0$, so only the second term on the RHS survives.  The requirement $(\mathcal{L}_X \Gamma)^i_{ij} = \partial_j \partial_k X^i = 0$ reduces the infinite dimensional symmetry group of the Carroll metric to a finite group.

Now the black hole event horizon is a null surface.  So local patches of the event horizon look like Carroll space \eqref{eq:carroll}.  For example, in ingoing Eddington-Finkelstein coordinates, the metric on the event horizon of a Schwarzschild black hole is
\beq\label{eq:horizon}
ds^2 = 0\cdot dv^2 + r^2 (d\theta^2 + \sin^2\theta d\phi^2).
\eeq
In other words, the black hole has an emergent, local Carroll invariance near the event horizon.  Price and Thorne's observation that the velocity of the horizon fluid is zero can be reinterpreted as a consequence of the fact that the horizon is a Carroll manifold (local patches look like Carroll space) and, in particular, the local speed of light is zero.

\subsection{Love numbers}

Finally, we observe that the above discussion is related to the vanishing of black hole Love numbers, at least for Schwarzschild black holes in $d=4$ general relativity.  The latter have been computed by  Binnington and Poisson \cite{Binnington:2009bb}.  Happily, Price and Thorne and Binnington and Poisson use the same ingoing Eddington-Finkelstein gauge to describe the near-horizon region, so translating between the two is relatively straightforward.  In particular, the membrane momentum density can be read off the metric:
\beq
\Pi^H_a	%= -g_{va}/(2\lambda) 
		= \alpha_H^{-2} \kappa g_{va} =\frac{1}{8\pi}\alpha_H^{-1} \kappa v_a.
\eeq
So,
\beq\label{eq:fluidv}
v_a = 8\pi \alpha_H^{-1} g_{va}.
\eeq
Binnington and Poisson \cite{Binnington:2009bb}  studied tidal perturbations of a Schwarzschild black hole.  Let $\mathcal{E}^l_A$ and $\mathcal{B}^l_A$ be the $l$th multipole moments of the gravitoelectric and gravitomagnetic components of the applied tidal field.  The index, $A$, runs over the spatial directions of the horizon $(\theta,\phi)$.  The perturbed metric has
\begin{align}
g_{vA}  = -&\frac{2}{(l-1)(l+1)}r^{l+1}e_4(r)\mathcal{E}^{l}_A\notag\\
&+\frac{2}{3(l-1)}r^{l+1}b_4(r)\mathcal{B}_A^{(l)},
\end{align}
where
\begin{align}
e_4 &= A_4 - 2\frac{l+1}{l}k_{\rm el}(2M/r)^{2l+1}B_4\label{eq:poisson1},\\
b_4 &= A_4 -2\frac{l+1}{l}k_{\rm mag}(2M/r)^{2l+1}B_4.\label{eq:poisson2}
\end{align}
The functions $A_4$ and $B_4$ are computed in \cite{Binnington:2009bb}.  $A_4$ represents the external tidal field and it vanishes at the event horizon (in fact, $A_4 = O(\alpha_H^2)$ so its contribution to \eqref{eq:fluidv} vanishes at the horizon).  The function $B_4$ represents the black hole's response.  It is nonzero (in fact, divergent) at the horizon.  The Love numbers, $k_{\rm el}$ and $k_{\rm mag}$, enter as coefficients multiplying $B_4$.  Comparing \eqref{eq:poisson1}-\eqref{eq:poisson2} with \eqref{eq:fluidv}, we see that vanishing of the fluid velocity, $v_a=0$, is equivalent to vanishing of the Love numbers.

\section{Discussion}
\label{sec:discuss}

We began by revisiting an old observation of Price and Thorne: the velocity of the membrane paradigm fluid velocity is zero and, as a result, the membrane does not respond to tidal forces in the usual way.  We related this observation to an 
emergent, local Carroll symmetry acting on the horizon.  The Carroll group is a subgroup of the BMS group and it would be interesting to understand if the full BMS group has a role to play in this story. 

We noted that for $d=4$ Schwarzschild black holes, the vanishing of the fluid velocity is related to vanishing of the black hole's Love numbers.  This hints at a connection between Carroll symmetry and black hole Love numbers.  However, black hole Love numbers are nonzero in higher spacetime dimensions \cite{Kol:2011vg}, in asymptotically anti de Sitter spacetimes \cite{Emparan:2017qxd}, and in modified theories of gravity \cite{Cardoso:2018ptl}.  So the possibility of an explanation for black hole Love numbers based on Carroll symmetry requires further investigation.

The Carroll symmetry we described acts on the event horizon.  In the EFT of black hole tidal interactions, the black holes are treated as point particles.  We expect some remnant of the Carroll symmetry to act on the fields of the EFT.  It would be very interesting to find this incarnation of the symmetry.

\hspace{0.2cm}

\begin{acknowledgments}
{\bf Acknowledgments} We are grateful to Miguel Campiglia, Laura Donnay, Lam Hui, Austin Joyce, Ira Rothstein, Luca Santoni, and Sam Wong for helpful discussions and to Roberto Emparan for comments on an earlier draft. This work is supported by Simons Foundation Award Number 555117.
 \end{acknowledgments}

\bibliography{ms.bib}

\end{document}